\documentclass[superscriptaddress,twocolumn,showpacs,preprintnumbers,amsmath,amssymb]{revtex4}
\usepackage{graphicx}
\usepackage{dcolumn}
\usepackage{bm}

\begin{document}

\title{Why we should teach the {B}ohr model and how to teach it effectively}

\pacs{01.40.Fk,01.40.G-,01.40.gb,01.50.ht}
\keywords{modern physics, atomic models, Bohr model}

\author{S. B. McKagan}
\affiliation{JILA and NIST, University of Colorado, Boulder, CO 80309, USA}

\author{K. K. Perkins}
\affiliation{Department of Physics, University of Colorado, Boulder, CO 80309, USA}

\author{C. E. Wieman}
\affiliation{Department of Physics, University of British Columbia, Vancouver, BC V6T 1Z1, CANADA}
\affiliation{JILA and NIST, University of Colorado, Boulder, CO 80309, USA}
\affiliation{Department of Physics, University of Colorado, Boulder, CO 80309, USA}

\date{February 29, 2008}

\begin{abstract}
Some education researchers have claimed that we should not teach the Bohr model of the atom because it inhibits students' ability to learn the true quantum nature of electrons in atoms.  Although the evidence for this claim is weak, many have accepted it.  This claim has implications for how to present atoms in classes ranging from elementary school to graduate school.  We present results from a study designed to test this claim by developing a curriculum on models of the atom, including the Bohr and Schr{\"o}dinger models.  We examine student descriptions of atoms on final exams in transformed modern physics classes using various versions of this curriculum.  We find that if the curriculum does not include sufficient connections between different models, many students still have a Bohr-like view of atoms, rather than a more accurate Schr{\"o}dinger model.  However, with an improved curriculum designed to develop model-building skills and with better integration between different models, it is possible to get most students to describe atoms using the Schr{\"o}dinger model.  In comparing our results with previous research, we find that comparing and contrasting different models is a key feature of a curriculum that helps students move beyond the Bohr model and adopt Schr{\"o}dinger's view of the atom.  We find that understanding the reasons for the development of models is much more difficult for students than understanding the features of the models.  We also present interactive computer simulations designed to help students build models of the atom more effectively.
\end{abstract}

\maketitle

\section{Introduction}

Atomic models are an important part of physics instruction at many different levels.  Students learn simple models of atoms in elementary school, and gradually integrate more complex ideas into these simple models throughout their education.  In physics graduate school, students continue to learn more sophisticated approaches to thinking about simple atoms.

Why is there such a focus throughout our education on atoms?  The study of atoms is a rich content area, providing a solid basis for understanding everything from the fundamental building blocks of nature to the basis of modern technology.  The structure of atoms is both beautiful and useful.  Furthermore, atomic models provide a good context for teaching scientific reasoning skills such as model-building and making inferences from observations.  The history of atomic models over the last century provides an exciting detective story in which students can be led through a complex web of reasoning about how new models are built and old models are discarded, based on a few simple observations.

\section{Previous Research and Controversy on student learning of atomic models}

Because quantum models of atoms are usually taught only at an advanced college level in the United States, there has not been much research on student learning of quantum models of atoms in this country.  Much more has been done in Europe, Asia, and Latin America, where quantum mechanics is often taught at the secondary level.~\cite{Fischler1992a,Fischler1992b,Blanco1998a,Justi2000a,Budde2002a,Budde2002b,Muller2002a,Kalkanis2003a,Petri1998a,Ke2005a,Harrison2000a,Taber2001a}  Within this literature, there is a great deal of controversy over whether to focus on the current scientific understanding~\cite{Fischler1992a,Fischler1992b} or to use a historical approach~\cite{Blanco1998a,Justi2000a}; whether to emphasize the commonalities~\cite{Budde2002a,Budde2002b} or the differences~\cite{Muller2002a,Kalkanis2003a} between classical and quantum physics; and which models are most appropriate for teaching students.

While much of the research literature emphasizes different models of the atom, different authors define and categorize models in different ways, with some explicitly focusing on historical models~\cite{Blanco1998a,Justi2000a}, some explicitly focusing on student models~\cite{Petri1998a}, some failing to distinguish between the two~\cite{Ke2005a,Harrison2000a}, and some explicitly creating hybrid models for teaching~\cite{Budde2002a,Budde2002b}.  Many of the authors do not explicitly define the different models they use, and each author appears to have different implicit assumptions about the ``correct'' model of the atom.  For example, Fischler and Lichtfeld~\cite{Fischler1992a,Fischler1992b} assume that the only ``correct'' explanation for the stability of atoms is localization energy due to the Heisenberg Uncertainty Principle; Justi and Gilbert~\cite{Justi2000a} assume that what we call the deBroglie model is an inappropriate blending of an orbit model and a quantum mechanical model; Harrison and Treagust~\cite{Harrison2000a} assume that it is correct to describe an electron cloud as ``made up of electrons that are moving very fast around the nucleus,'' and that a ``cloud'' model is more sophisticated than a ``shell'' model (without defining either of these terms).

There are many definitions of ``models'' in the literature.  In this paper, we will focus on scientific models, as defined by Halloun~\cite{Halloun2004a}: ``A scientific model is... a conceptual system mapped, within the context of a specific theory, onto a specific pattern in the structure and/or behavior of a set of physical systems so as to reliably represent the pattern in question and serve specific functions in its regard.''  Another useful definition is given in the U.S. National Science Education Standards~\cite{NRC1996a}: ``Models are tentative schemes or structures that correspond to real objects, events, or classes of events, and that have explanatory power. Models help scientists and engineers understand how things work.''  We note that both of these definitions focus on the usefulness, rather than the correctness, of a model, so that different models may be appropriate in different contexts.

Another common, and very different, use of the term ``model'' is to refer to students' mental models, which may or may not match accepted scientific models~\cite{Gentner1983a}.  This is a controversial use of the term, as there is evidence that student thinking is often fragmented and context-dependent~\cite{diSessa1993a,Elby2001a}, and stating that students have mental models implies more coherence to their thinking than may be warranted.

While there is controversy over the structure of students' thinking while they are in the process of learning, there appears to be broad consensus that expert scientists do indeed use models, and one goal of science education is to help students learn to use models in the same way.  According to the recent National Research Council report \emph{Taking Science to School}~\cite{NRC2007a}, science is ``both a body of knowledge and an evidence-based, model-building enterprise that continually extends, refines, and revises knowledge.''  Thus, we do not make any claims about whether students, in the process of learning, use mental models or not.  Rather, we attempt to measure, at the end of a course on modern physics, to what degree students are expressing various expert scientific models.

In this paper, we will focus on three historical models of the Hydrogen atom, as proposed by Bohr~\cite{Bohr1913a}, deBroglie~\cite{deBroglie1924a}, and Schr{\"o}dinger~\cite{Schrodinger1926a}.  In the Bohr model, electrons are point particles that move around the nucleus in circular orbits at fixed radii.  In the deBroglie model, electrons are standing waves on rings with the same radii as the Bohr model.  In the Schr{\"o}dinger model, electrons are clouds of probability whose density is given by the solutions to the three-dimensional Schr{\"o}dinger equation for the Coulomb potential that the electron feels from the nucleus.

One of the most controversial questions in teaching models of the atom is whether and how to teach the Bohr model.  This is an area of active debate among high school teachers, as evidenced by a recent discussion among a group of local high school teachers over district standards, in which teachers could not agree over which model to teach.~\cite{Loeblein2007a}  The resulting standards state that students should be able to ``Describe the properties and relationships of neutrons, protons, and electrons,'' but do not state which model should be used to describe electrons.~\cite{Jeffco2005a}  Likewise, the U.S. National Science Education Standards state that ``Each atom has a positively charged nucleus surrounded by negatively charged electrons,'' but do not describe the properties of those electrons.~\cite{NRC1996a} Thus, both national and district standards are silent on the question of which model(s) should be used to describe atoms.

Within the education research community, at one extreme, Taber~\cite{Taber2001a} states, ``Clearly a full understanding of the modern notion of the atom based on quantum mechanics would \textit{not} be appropriate at secondary level,'' and advocates sticking to the Bohr model.  Most researchers in this area advocate teaching a quantum mechanical view of atoms at the secondary level and beyond, but disagree about whether the Bohr model is an appropriate step towards this path.  Petri and Niedderer~\cite{Petri1998a} view a Bohr-like model as a necessary step in the learning pathway of a student.  Those who advocate a historical approach view the Bohr model as an important historical step in understanding atoms~\cite{Blanco1998a,Justi2000a}.  Those who believe that learning quantum mechanics requires contrasting it with classical intuition view the Bohr model as a useful tool in refining the quantum mechanical view of atoms~\cite{Muller2002a,Kalkanis2003a}.  At the opposite extreme, Fischler and Lichtfeld~\cite{Fischler1992a,Fischler1992b} claim that the Bohr model is an obstacle to learning the true quantum nature of atoms, and state, ``In the treatment of the hydrogen atom, the model of Bohr should be avoided.''

The work of Fischler and Lichtfeld is often cited in the Physics Education Research (PER) community and beyond as evidence that it is preferable to avoid teaching the Bohr model entirely~\cite{Zollman2002a,Fletcher2004a,Ireson2000a}, and their claims have been incorporated into curriculum design~\cite{Zollman2002a}.  Because the implications of these claims are so far-reaching, it is worth taking a closer look at the research behind them.

In our view, the work of Fischler and Lichtfeld does not provide convincing evidence that teaching the Bohr model prevents students from learning the Schr{\"o}dinger model.  In Refs. [1-2], they discuss a high school quantum mechanics curriculum they have developed, known as the ``Berlin Concept of Quantum Physics.''  They implemented this curriculum in 11 courses (the test group), and took extensive data in these courses as well as in 14 other courses implementing a more standard quantum mechanics curriculum (the control group).  They do not describe what the standard quantum mechanics curriculum is like or how it is different from their curriculum, so it is difficult to determine whether any of the results they show are due to the avoidance of the Bohr model or to other differences in the curriculum.

In Ref. \cite{Fischler1992a} they compare the network of ideas before and after instruction for one student in the test group and one student in the control group, and show that the student in the test group is thinking more quantum mechanically than the student in the control group.  In Ref. \cite{Fischler1992b}, they present data from about 100 students in each group, showing that when they asked the students to explain why the atom is stable, many more students in the test group ($68\%$) than in the control group ($7\%$) talked about localization and the Heisenberg uncertainty principle.  They also show a graph indicating the percentage of students in each group whose ``conceptual change'' over the course of instruction was none, little, satisfactory, or complete, indicating that more students in the test group ($67\%$) than in the control group ($2\%$) had satisfactory or complete conceptual changes.  There is no explanation of what data this graph is based on or how they decided what category a student's conceptual change falls into or even what they mean by conceptual change.

Fischler and Lichtfeld assume that the only goal of instruction on atoms is for students to acquire the content knowledge of the Schr{\"o}dinger model.  If one follows their advice, it is not possible to use atoms as a context for model-building, which is one of our main goals in teaching atomic models.

Further, rather than making exclusive use of one expert model, real experts are able to use multiple models simultaneously, recognizing the strengths and limitations of each one and applying them appropriately.~\cite{Brookes2006b,Brookes2007a,Grosslight1991a} When solving simple problems, practicing scientists often use the Bohr model.  Therefore, avoiding the Bohr model deprives students of a tool that scientists find useful.

Finally, research shows that avoiding discussion of topics likely to lead to misconceptions doesn't work.~\cite{Bransford1999a}  It is much more effective to explicitly address the problems students are likely to encounter.  This approach is especially important when discussing instruction on atomic models at the university level.  Students learn aspects of the Bohr model in high school, in elementary school, and in popular culture, so they enter university courses with preconceived ideas about atoms, whether we like it or not.

\section{The Study}
Our basic research question is: ``Is teaching the Bohr model an obstacle to learning the Schr{\"o}dinger model of the atom?''  This is an experimental question that is relatively straightforward to answer with a simple study.  The ideal study to answer this question would be to compare student conceptions of atoms in two courses, one where the Bohr model is taught and one where it is avoided.  We were not able to conduct this ideal study because we could not convince any faculty teaching modern physics at the University of Colorado to not teach the Bohr model.  However, to show that the Bohr model is \textit{not} an obstacle to learning the Schr{\"o}dinger model, it is necessary only to show that in a well-designed course that includes the Bohr model, most students eventually develop an understanding of atoms based on the Schr{\"o}dinger model.  This, therefore, is what we set out to do.

The context of the study is a transformed modern physics course for engineering majors developed and taught by the authors for two semesters (Fa05 and Sp06) and then taught by another professor in the PER group for the following two semesters (Fa06 and Sp07), using the same materials (with exceptions discussed below).~\cite{McKagan2007a}  The course format was a large lecture, with an enrollment of 189, 184, 94, and 153 in Fa05 through Sp07, respectively.  The content of the course emphasized reasoning development, model building, and connections to real world applications.  In addition we implemented a variety of PER-based learning techniques, including concept tests, peer instruction, collaborative homework sessions, and interactive simulations.  The curriculum is available online.~\cite{2130}

\begin{figure}[htbp]
  \includegraphics[width=\columnwidth]{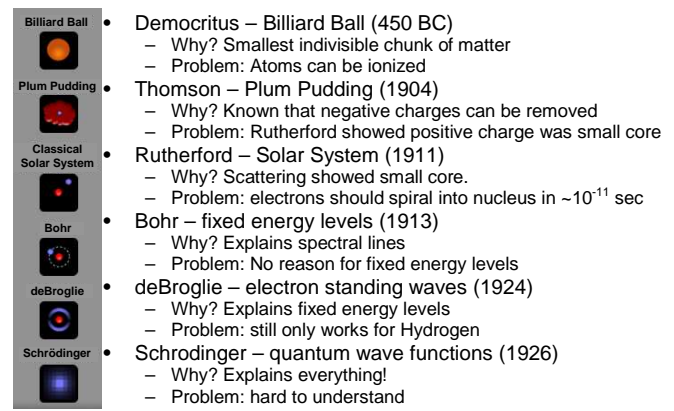}
  \caption{Models of the Atom discussed in class}
  \label{models}
\end{figure}

The treatment of atomic models in this course is outlined in Fig.~\ref{models}.  In lectures we discussed the historical development of different models, with a focus on model-building and the reasoning for each model.  In moving to each new model, we emphasized the shortcomings of the previous model and asked students to consider how to resolve these shortcomings.  The homework included many questions asking students to compare and contrast models and to discuss the advantages and limitations of each.  While we did work out the mathematics of the Bohr model in lecture, the main emphasis was on the reasoning behind it, rather than on calculations.  We did not solve the Schr{\"o}dinger equation for the Hydrogen atom, but discussed how it is solved and the properties of the solutions.

All four semesters, the final exam included the following question (Students were not allowed to keep copies of the final exams, and solutions were never provided.):
\begin{quote}
A hydrogen atom is in its lowest energy state.  Use words, graphs, and diagrams to describe the structure of a Hydrogen atom \textbf{in its lowest energy state (ground state)}. Include in your description:
\begin{itemize}
  \setlength{\itemsep}{1pt}
  \setlength{\parskip}{0pt}
  \setlength{\parsep}{0pt}
    \item At least two ideas that are important to any accurate description of a hydrogen atom.
    \item An electron energy level diagram of this atom, including numerical values for the first few energy
levels, and indicating the level that the electron is in when it is in its ground state.
    \item A diagram illustrating how to accurately think about the distance of the electron from the nucleus for
this atom.
\end{itemize}
(On these diagrams, be quantitative where possible. Label the axes and include any specific information
that can help to characterize hydrogen and its electron in this ground state.)
\end{quote}

We designed this question to measure which models students naturally use to think about atoms.  The third bullet point, by asking students to describe how to ``think about the distance of the electron from the nucleus,'' forces them to distinguish between the Bohr and Schrodinger models, since the electron is at a fixed distance in the Bohr model, and spread over a range of distances in the Schrodinger model.  We asked the question in a way that does not explicitly mention models, so that it would not directly prompt the students to reflect on which models they are using to describe the atom.  This decision was based on research showing that asking students to engage on metacognitive reflections of their own thought process can cause them to think more deeply and change their answer~\cite{Berardi-Coletta1995a}.  Thus, if we had asked them to reflect on models of the atom, many students would be likely to use the Schrodinger model who would not otherwise use it.

In Fa05, the first semester we taught the course, this question was followed by a second part asking the students to describe the effect of increasing the angular quantum number $\ell$ by $1$.  Because answering this second part requires using the Schr{\"o}dinger model, we suspected that it might prompt students to use the Schr{\"o}dinger model in the first part more than they might otherwise.  As will be discussed in Section IV, this was indeed the case.  We did not ask this second part in any of the remaining three semesters.

The criteria we used to evaluate the answers are given in Section IV, and examples are given in Appendix A.

\section{Results}
We analyzed responses for all students who attempted to answer the final exam question described above in each semester of the course.  We determined the models that students were using with the following coding scheme:

\begin{description}
  \setlength{\itemsep}{1pt}
  \setlength{\parskip}{0pt}
  \setlength{\parsep}{0pt}
    \item[Bohr] Explicitly mentions Bohr model, draws or describes orbits, or says or implies that electron is at fixed radius
    \item[deBroglie] Explicitly mentions deBroglie model, or draws or describes standing wave around ring
    \item[Schr{\"o}dinger] Explicitly mentions Schr{\"o}dinger model, draws or describes electron cloud, or draws or describes probability distribution.
\end{description}

For each of the three models, we noted whether the student used the model implicitly, by describing one of its characteristics but not naming the model, or explicitly, by naming the model.  We also noted when students pointed out the incorrectness or limitations of a model, either implicitly, by criticizing some characteristic of the model, or explicitly, by naming the model itself as incorrect or limited.

We did not attempt to characterize whether the students were applying the models correctly.  Most student statements could be construed as correct according to some model.  Appendix A gives sample student responses and further discussion of how they were coded, including testing for inter-rater reliability.

\begin{figure}[htbp]
  \includegraphics[width=\columnwidth]{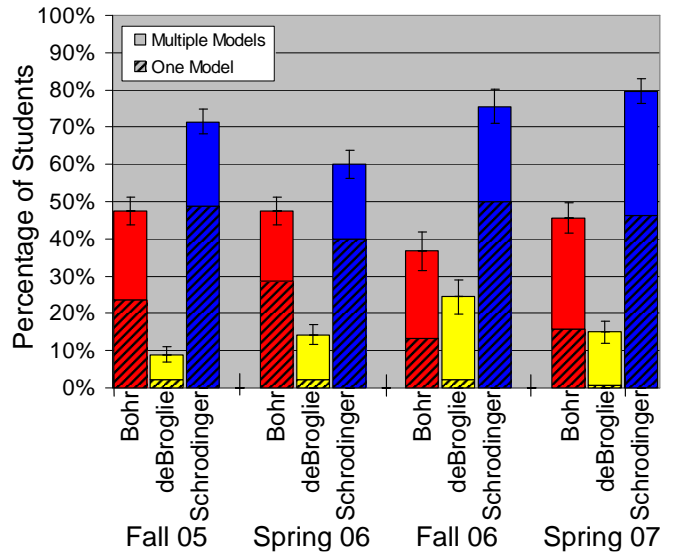}
  \caption{Models used to describe Hydrogen atom (either implicitly or explicitly).  The lower section of the bar for each model indicates the percentage of students who used that model exclusively, while the upper section indicates the percentage of students who used that model along with at least one of the other two models.  The total height of each bar thus indicates the total percentage of students who used each model, whether exclusively or not.  These percentages include only students who discussed the corresponding models as valid and without limitations.  Error bars represent the standard error on the mean for the total height of each bar.  N = 179, 175, 90, 147 for each of the respective semesters.}
  \label{data1}
\end{figure}

\begin{figure}[htbp]
  \includegraphics[width=\columnwidth]{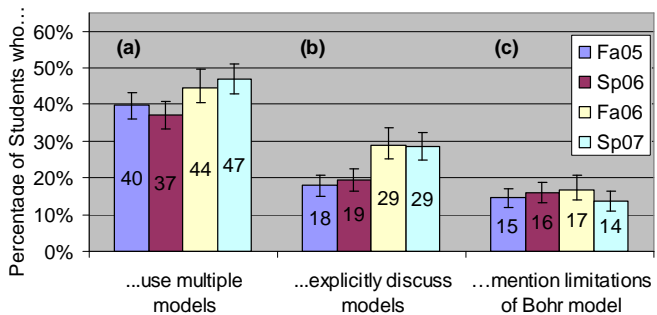}
  \caption{Percentage of students (a) using multiple models, (b) explicitly discussing models, and (c) mentioning limitations of Bohr model. (a) includes students who used multiple models both implicitly and explicitly.  (b) includes both students who explicitly referred to models without criticizing them, and students who explicitly discussed the limitations or incorrectness of a model or models. (c) includes students who mentioned the limitations or incorrectness of the Bohr model or some aspect of it.  Error bars represent the standard error on the mean.}
  \label{summary}
\end{figure}

Fig.~\ref{data1} shows the percentage of students using each model in each of the four semesters.  We note that while many previous studies have treated student models as mutually exclusive, so that each student has only one model of the atom, we found it impossible to code our student responses in this way.  As can be seen in Fig.~\ref{summary}a, a large fraction of students used multiple models, either by implicitly blending ideas from different models or by explicitly comparing and contrasting models.

Our goal is for all students to use the Schr{\"o}dinger model, but not necessarily to the exclusion of other models.  We view it as desirable for students to use multiple models, especially if they do so explicitly, as this is closer to the practice of expert physicists.  However, we recognize that student use of multiple models is not always positive, since students sometimes simply confuse the models, rather than knowing how to apply each one in the appropriate context, as experts do.  Because we did not ask students to apply the models, it is difficult to judge whether their use of multiple models was expert-like or not.  In some cases, it certainly was, as the students explicitly discussed the limitations of each model.  In other cases, it was not clear, as the students did not give enough detail for us to judge how they viewed the models.  However, we suspect that many experts answering our exam question would also fail to be explicit about their use of models, although they would be able to distinguish the models if asked to do so.  Since we cannot know whether students are failing to explicitly discuss models because they do not know how or because they don't think it is important in answering this question, we cannot fault them.

After reading students' responses to the exam question from the first two semesters, we were disappointed by how Bohr-like the students' models were.  While a majority of the students used the Schr{\"o}dinger model (height of first two blue bars), approximately a quarter ($23\%$ and $29\%$ in the respective semesters) were using the Bohr model alone (height of hashed part of first two red bars).  Furthermore, after removing the second part of the exam question in Sp06, the percentage using the Schrodinger model dropped significantly ($p=0.01$) from $72\%$ to $60\%$.  Since there are no other significant differences between the types of student responses in the two semesters, we believe this drop is because the second part of the exam question prompted them to use the Schr{\"o}dinger model.

After reviewing our curriculum, we realized that the treatment of atomic models was too disjointed.  We spent a considerable amount of time at the beginning of the course discussing the evolution of models through deBroglie, and then spent about five weeks on electron waves and the Schr{\"o}dinger equation and its applications, before returning to the Schr{\"o}dinger model of the atom.  The material in between was necessary background for understanding the Schr{\"o}dinger model.  However, when we did get to the Schr{\"o}dinger model of the atom, we did not spend much time relating it back to the previous models of the atom.  Based on interviews and observations of students doing homework during the first two semesters, we suspect that many of them did not see any connection between the Schr{\"o}dinger model and the earlier models of the atom.  Instead, they viewed this model as simply one more example of an application of the Schr{\"o}dinger equation in a different potential.

Starting in Fa06, we changed the treatment of the Schr{\"o}dinger model to focus on relating it back to the previous models and recognizing why it was an improvement.  In addition, we put a greater emphasis on model-building throughout the course, using a touchstone example developed by Dykstra~\cite{Dykstra2006a} of a farmer attempting to discover the underlying structure of his seeds by examining his sprouts.  In Sp07, we started using the interactive simulations \textit{Rutherford Scattering} and \textit{Models of the Hydrogen Atom} (see Appendix B for details; both simulations are available for free download from the Physics Education Technology (PhET) website~\cite{PhET}), along with a detailed homework exercise in which students used the simulation to analyze the differences in the experimental predictions of each model and explain the reasoning behind the development of the models.

After changing the curriculum to address the problems seen in the first two semesters, we found that $76-80\%$ of students were using the Schr{\"o}dinger model (height of last two blue bars - significantly lower than Sp06 ($p<0.05$\footnote{All statistics were calculated using a one-tailed Z test under the hypothesis that the treatment would lead to a decrease in the exclusive use of the Bohr model and an increase in the use of the Schrodinger model, use of multiple models, explicit discussion of models, and mention of the limitations of the Bohr model.}) and comparable to Fa05, but without the prompting of the second part of the question), and only $13-16\%$ were using the Bohr model alone (height of hashed part of last two red bars - significantly less than either of the first two semesters ($p<0.05$)).  These results suggest that teaching the Bohr model does not prevent students from learning the Schr{\"o}dinger model, and that the improvements to the curriculum made in Fa06 improved student learning of the Schr{\"o}dinger model.

Further, as can be seen in Fig.~\ref{summary}, the percentage of students explicitly discussing models increased after the further emphasis on model-building and the connection between models was added in Fa06 ($p<0.05$).

A possible concern about comparing the results from different semesters in Figs.~\ref{data1} and~\ref{summary} is that in addition to different curricula in Fa06 and Sp07, there was a different instructor.  However, the new instructor worked closely with one of the old instructors to implement the course in a similar way, and in other areas of the course where the curriculum did not change, there was no significant difference in student responses between Sp06 and Fa06 (for example, see Ref. \cite{McKagan2008c}).  Further, there were no statistically significant differences between any of the four semesters on any of the 23 multiple choice questions that were asked on all four final exams (out of a total of 38, 38, 35, and 33 multiple choice questions in each of the respective semesters).

Another possible concern is that this exam question was one of 3-5 long answer questions on the final exam in Fa05 and Sp06, but was the only long answer question on the exam in Fa06 and one of only two in Sp07, so students had more time to work on it.  This was reflected in the more detailed answers that students gave in Fa06 and Sp07, and it would not be fair to compare the quality of the explanations between these semesters.  However, because the question does not ask the students to reflect on the nature of models, giving them more time to think about it should not lead them to use different models.  To eliminate any possible effects of exam length, our study excluded students who left the exam question blank (4, 2, zero, and zero students in the respective semesters).

\section{Conclusion and Next Steps}
This study has answered the original research question, ``Is teaching the Bohr model an obstacle to learning the Schr{\"o}dinger model of the atom?'' in the negative.

After completing our study, we discovered two previous studies in radically different contexts, one in a German secondary school course~\cite{Muller2002a}, and the other in a Greek course for prospective/in-service teachers~\cite{Kalkanis2003a}, that found similar results to ours.  In both studies, researchers introduced a transformed modern physics curriculum with an emphasis on contrasting models, including the Bohr model.  In both cases, when students were asked at the end of the course to draw a picture of a Hydrogen atom, most ($77\%$ in the first study and $99\%$ in the second study) drew a quantum model (as determined by the researchers).

While neither of these studies focused on our research question, both support the conclusion that teaching the Bohr model is not an obstacle to learning the Schr{\"o}dinger model.  In comparing our curriculum with the curricula used in these previous studies, one key feature stands out: all three emphasized comparing and contrasting the Bohr model with later models.  It is likely that this key feature is a necessary component for a curriculum to help students put the Bohr model in proper context and move beyond it.  This conclusion is consistent with other research showing that students can learn much more about a case by comparing and contrasting it with other cases than by studying a single case alone.~\cite{Loewenstein2003a,Marton2006a,Bransford1999b}

This study has opened up many questions for further research.  In interviews and observations of students doing homework, we have found that while many students were able to correctly answer homework questions about atomic models and describe the salient features of each model, they seemed to view each model as a list of characteristics to be memorized, rather than having a clear mental picture of the model.  This is in sharp contrast to our observations of students describing other topics in the course for which we used interactive simulations~\cite{McKagan2008b}, such as lasers, discharge lamps, and the photoelectric effect.  For these topics, students were able to give vivid and detailed descriptions and apply their knowledge to new situations without apparent effort.

We have found throughout the course that students did not develop model-building skills to the degree we hoped.  Most students could describe the models, but had trouble giving satisfactory reasoning for why each model surpassed the previous one.  When we asked them to make inferences from observations, they often confused the two, and did not give a logical argument from one to the other.  Research by Etkina et al.~\cite{Etkina2006a} suggests that scientific reasoning skills are difficult to develop without a curriculum specifically aimed at giving students practice engaging in scientific model building.  While several such curricula have been developed for introductory physics \cite{Etkina2001a,McDermott1996a,Goldberg2005a}, little work in this area has been done in modern physics.

In response to these problems, we have designed the two new interactive computer simulations discussed in Appendix B. As discussed in the previous section, we integrated the simulations into the course in Sp07 and students used them to work through the reasoning behind the development of each model. While the use of the simulations and associated homework did not increase percentage of students using the Schr{\"o}dinger model (Fig.~\ref{data1}), the percentage of students using multiple models and explicitly discussing models did increase significantly after introducing the simulation in Sp07 (Fig.~\ref{summary}).  Further, observations of students working through simulation in homework sessions indicate that they were engaging in model-building to a degree that they had not in other aspects of the course, including previous homework on atomic models.  Further research is needed to determine whether and how the simulation can be used most effectively to help students engage in model building.

\section{Acknowledgments}
We thank Noah Finkelstein for allowing us to work with his course in Fa06 and Sp07, and Charles Baily for assistance with coding, helpful discussions, and feedback.  We thank Chris Malley, the software engineer for the \textit{Rutherford Scattering} and \textit{Models of the Hydrogen Atom} simulations.  We also thank the PhET team and the Physics Education Research Group at the University of Colorado.  This work was supported by the NSF, The Hewlett Foundation, and the University of Colorado.

\appendix
\section{Sample responses}
\begin{figure}[htbp]
  \includegraphics[width=\columnwidth]{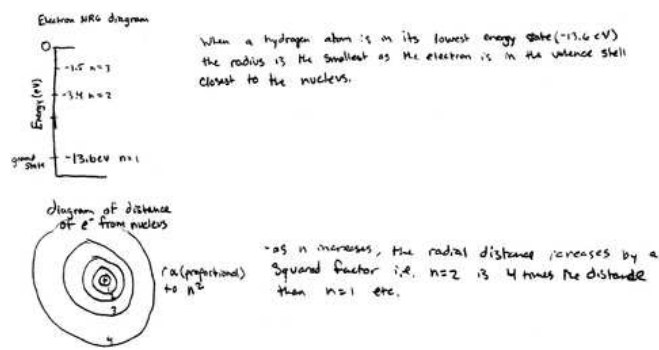}
  \caption{Example of a student response coded as implicitly using the Bohr model only.  This student described the electron as if it had a fixed radius, and drew orbits of fixed radii, a characteristic of the Bohr model.}
  \label{bi}
\end{figure}

\begin{figure}[htbp]
  \includegraphics[width=\columnwidth]{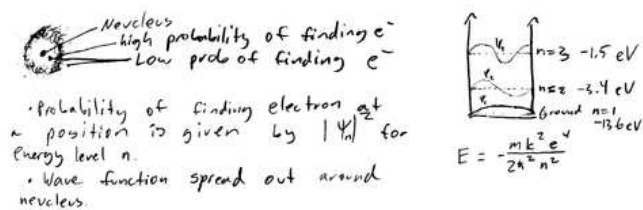}
  \caption{Example of a student response coded as implicitly using the Schr{\"o}dinger model only.  This student drew an electron cloud and described how it characterized the probability of finding the electron at different locations, a characteristic of the Schr{\"o}dinger model.}
  \label{si}
\end{figure}

\begin{figure}[htbp]
  \includegraphics[width=\columnwidth]{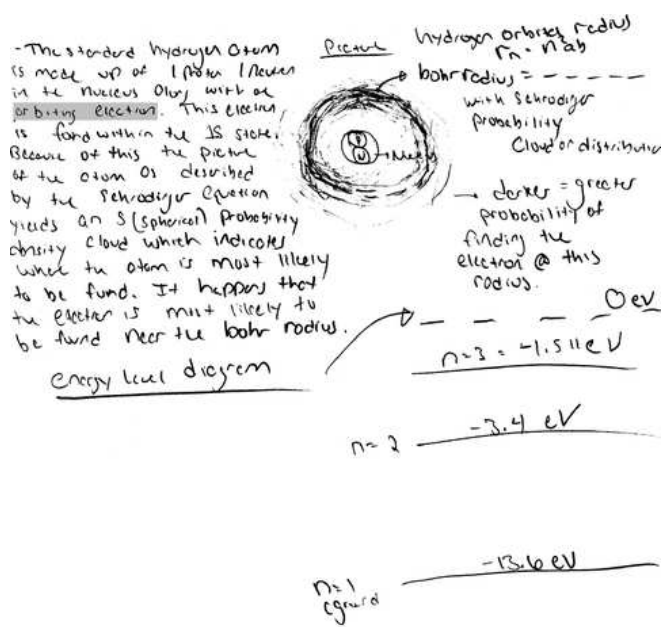}
  \caption{Example of a student response coded as implicitly using the Bohr and Schr{\"o}dinger models.  This student drew and described a probability cloud or distribution, a characteristic of the Schr{\"o}dinger model, but also described an ``orbiting electron'' (highlighted), a characteristic of the Bohr model.}
  \label{bisi}
\end{figure}

\begin{figure}[htbp]
  \includegraphics[width=\columnwidth]{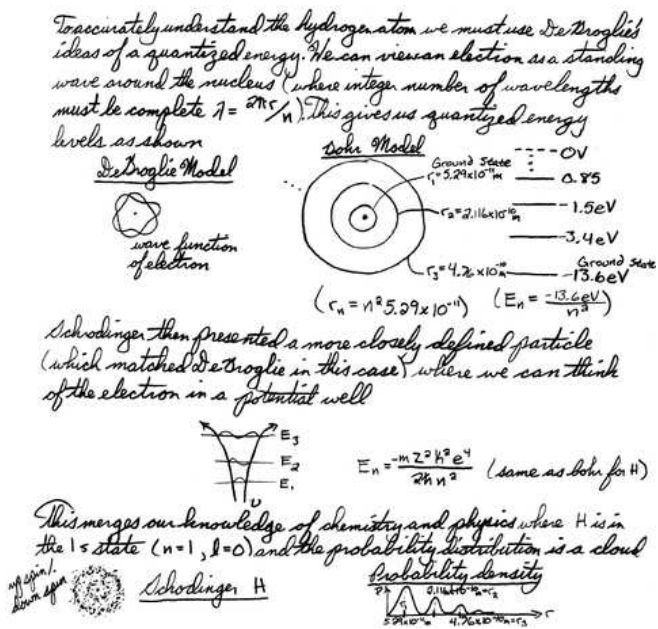}
  \caption{Example of a student response coded as \textit{explicitly} using the Bohr, deBroglie, and Schr{\"o}dinger models.  This student described the characteristics of each model, explicitly labeling them as ``Bohr model,'' ``deBroglie model,'' and ``Schodinger H.''}
  \label{bedese}
\end{figure}

\begin{figure}[htbp]
  \includegraphics[width=\columnwidth]{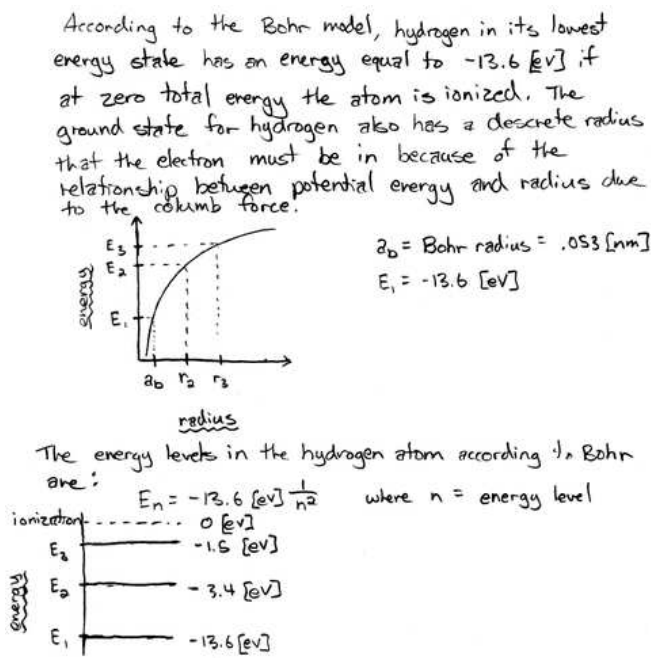}
  \caption{Example of a student response coded as \textit{explicitly} using the Bohr model only.  This student began the description with ``According to the Bohr model,'' and described the electron as being at a discrete radius, a characteristic of the Bohr model.}
  \label{be}
\end{figure}

Figs. \ref{bi}-\ref{be} show sample student responses to illustrate the application of our coding scheme.  As there is some confusion in the literature between historical models and student models, we note that by coding a student response as using a particular historical model, we do not mean to imply that the student is using the model as it was understood historically by its creator.  We mean only that the student is using some vital element of the model as we have defined it in Section II.

Some previous literature makes much finer distinctions in student thinking than the three categories we have used, for example, distinguishing between the conceptions of smeared orbits, wave functions that don't become real until you measure it, and electron clouds~\cite{Petri1998a}, or distinguishing between orbitals and electron clouds~\cite{Muller2002a}, all of which we characterize as being part of the Schrodinger model.  In practice, we have found that it is not possible to make such fine distinctions based on written responses alone.

We did not attempt to code for whether the students were using the models ``correctly,'' as correctness is difficult to define for this question.  While several students made minor errors such as stating that a Hydrogen atom nucleus contains a neutron (about $10\%$ of students), as in Fig.~\ref{bisi}, or drawing the probability density for an excited state rather than a ground state, as in Fig.~\ref{bedese}, most students correctly described some important features of some model.  Some student responses were more complete than others, but because the question was open-ended, students focused on different features of the models, and a more complete response does not necessarily imply a more complete understanding.

A sample of $50$ student responses from each semester were coded independently by one of the authors (SBM) and by another coder.  The second coder was not told about the purpose of the study or the differences between the curricula in the different semesters until after the coding was complete.  The initial inter-rater reliability was $83\%$.  The two coders refined the coding scheme by discussing and reaching consensus on a quarter of the responses where they disagreed.  They then recoded the remainder of the sample of responses independently with a new inter-rater reliability of $93\%$.  The remaining student responses were coded by SBM using the refined coding scheme.

Figs. \ref{bi}-\ref{bisi} show three examples of responses coded as implicitly, rather than explicitly, using models.  These students characterized atoms using characteristics that we identified as belonging to specific historical models, but the students did not mention these models by name.

Fig.~\ref{si} shows one of several responses for which the two coders disagreed about whether the student was also using the deBroglie model.  The drawing on the right could be interpreted as a spread out deBroglie wave, or as a one dimensional wave function for an infinite square well.  In the end, because it was ambiguous, this response and others like it were not coded as using the deBroglie model.

Figs. \ref{bedese}-\ref{be} show two examples of responses coded as explicitly using models.  References to the ``Schr{\"o}dinger equation'' or the ``Bohr energies'' or even references to what Bohr, deBroglie, or Schr{\"o}dinger discovered historically were not coded as explicit references to models unless the students were clearly comparing and contrasting different scientists' views or treating these views as models.

\section{Interactive Simulations}

\begin{figure}[htbp]
  \includegraphics[width=\columnwidth]{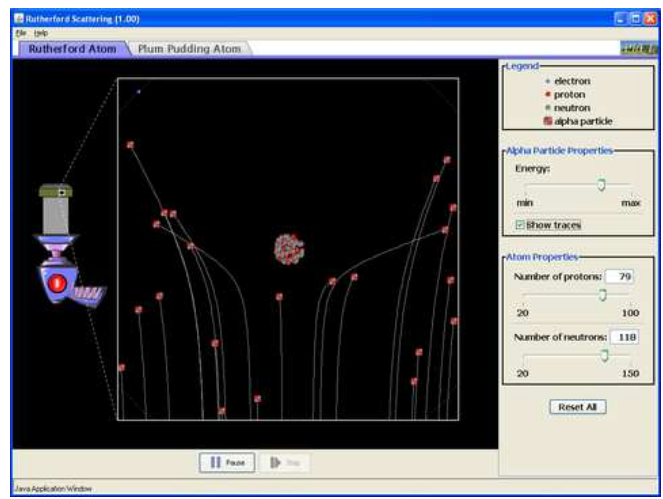}
  \caption{\textit{Rutherford Scattering} simulation}
  \label{rutherford}
\end{figure}

\begin{figure}[htbp]
  \includegraphics[width=\columnwidth]{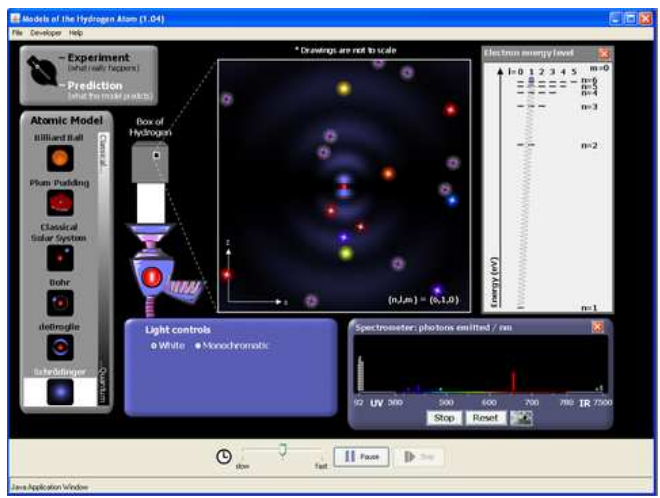}
  \caption{\textit{Models of the Hydrogen Atom} simulation}
  \label{MHAsim}
\end{figure}

In response to the problems we found in this study, we have designed two new interactive computer simulations called \textit{Rutherford Scattering} (see Fig.~\ref{rutherford}) and \textit{Models of the Hydrogen Atom} (see Fig.~\ref{MHAsim}).  These simulations, which were developed as part of the Physics Education Technology (PhET) project, can be downloaded for free from the PhET website~\cite{PhET}.  They are designed to be used as part of guided inquiry exercises to help students build models of the atom and develop an understanding of the reasoning behind each model.  The homework exercises we used in Sp07 are available from both the PhET activities database and from our modern physics course archive~\cite{2130}.

The \textit{Rutherford Scattering} simulation was designed in response to observations that after instruction on Rutherford Scattering, many students could not understand \emph{why} alpha particles scatter from atoms in the way they do, why the actual behavior is different from the predictions of Thomson's Plum Pudding model, and why this discrepancy is important.  In the simulation, students can see the microscopic effects of shooting alpha particles at a Plum Pudding atom and a Rutherford atom and build up a model for why the behavior of the alpha particles is different in these two cases.

The \textit{Models of the Hydrogen Atom} simulation was designed in response to observations that after instruction on models of the atom, many students did not have clear mental pictures of the different models, and did not understand the reasoning that led to the models.  In the simulation, students can see an animated visual representation of each model, see how shooting light at different models leads to different output on a spectrometer, and see the connection between the physical picture of the atom and the energy level diagram.

In response to a possible concern that by simulating the internal structure of atoms, we are ``giving away'' the model rather than making the students construct it for themselves~\cite{Heron2004a}, we note that in interviews and homework sessions, the students we observed using the simulation did not grasp the models being displayed immediately, but built their understanding slowly as they played with the simulation or worked through the homework.  Even with all the cues ``given'' in the simulation, the students we observed still appear to need to construct their own models before they can make sense of what they are seeing.

\begin{figure}[htbp]
  \includegraphics[width=\columnwidth]{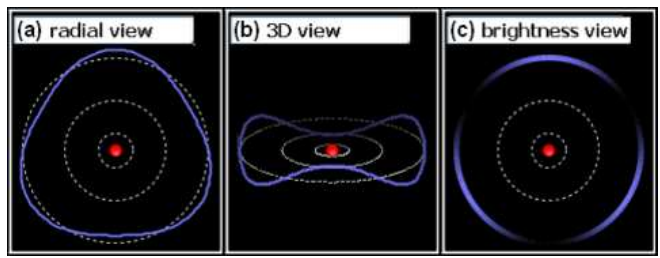}
  \caption{Three representations of the deBroglie model}
  \label{debroglie}
\end{figure}

Because the mental leap between the Bohr and Schr{\"o}dinger models is so large, we believe that the deBroglie model serves as a useful bridge, providing a link to both the fixed orbits of the Bohr model and the waves of the Schr{\"o}dinger model.  Even within the research literature that emphasizes comparing and contrasting models, the deBroglie model is often left out~\cite{Blanco1998a,Justi2000a,Budde2002a,Budde2002b,Muller2002a,Kalkanis2003a}.  We have observed that the idea of a three-dimensional wave that does not have a simple sinusoidal structure is quite difficult for many students, and as a result, they often have trouble recognizing the connection between the ``electron clouds'' of the Schr{\"o}dinger model, and the ``electron waves'' they have learned about in one dimension.

To help students make the connection between the more traditional representation of sinusoidal waves and the representation of brightness as the magnitude of probability density used for the the Schr{\"o}dinger model in Fig.~\ref{MHAsim}, the \textit{Models of the Hydrogen Atom} simulation provides several different representations of wave amplitude in the deBroglie model, as shown in Fig.~\ref{debroglie}.  The first two representations help students recognize the deBroglie waves as waves, and the third representation helps students connect these waves to the representation in the Schrodinger model.  In student interviews on the \textit{Models of the Hydrogen Atom} and \textit{Radio Waves and Electromagnetic Fields} simulations, we have found that students do not recognize the brightness representation shown in Fig.~\ref{debroglie}c as a wave if it is the only representation used, but can correctly interpret it after comparing it with other representations.

While further research is needed to carefully test the effects of the \textit{Rutherford Scattering} and \textit{Models of the Hydrogen Atom} simulations and associated activities on student reasoning about atomic models, initial observations indicate that these can be useful tools for helping students engage in model-building.

\bibliography{../bibliographies/PER}
\bibliographystyle{apsrev}
\end{document}